\documentclass[%
 reprint,
 amsmath,amssymb,
 aps,
prl,
floatfix,
]{revtex4-1}

\usepackage{graphicx}
\usepackage{subfigure}
\usepackage{dcolumn}
\usepackage{bm}

\begin{document}

\preprint{APS/123-QED}

\title{Orientational Effects on the Amplitude and Phase of Polarimeter Signals in Double Resonance Atomic Magnetometry}

\author{Stuart J. Ingleby}
 \email{stuart.ingleby@strath.ac.uk}
\author{Carolyn O'Dwyer}
\author{Paul F. Griffin}
\author{Aidan S. Arnold}
\author{Erling Riis}
\affiliation{%
 Department of Physics, SUPA, Strathclyde University, 107 Rottenrow East, Glasgow, UK\\
}%
\date{\today}

\begin{abstract}
Double resonance optically pumped magnetometry can be used to measure static magnetic fields with high sensitivity by detecting a resonant atomic spin response to a small oscillating field perturbation. Determination of the resonant frequency yields a scalar measurement of static field ($B_0$) magnitude. We present calculations and experimental data showing that the on-resonance polarimeter signal of light transmitted through an atomic vapour in arbitrarily oriented $B_0$ may be modelled by considering the evolution of alignment terms in atomic polarisation. We observe that the amplitude and phase of the magnetometer signal are highly dependent upon $B_0$ orientation, and present precise measurements of the distribution of these parameters over the full $4\pi$ solid angle. 
\end{abstract}

\maketitle

\section{Introduction}

Optically pumped magnetometers, exploiting sharp magnetic resonance features in optically pumped alkali metal vapours, can be used for high sensitivity measurement of magnetic fields \cite{Sheng2013,Schwindt2007}. The Larmor frequency $\omega_L$ of atomic spin precession in a polarised sample is proportional to the magnetic field magnitude $|B_0|$ and can be determined by measurement of a resonant response in atomic magnetisation to an oscillating perturbation of frequency $\omega_{RF}$. This perturbation can be applied through modulation of pump light amplitude \cite{Bell1961}, frequency \cite{Jimenez-Martinez2010} or polarisation \cite{Breschi2014}, or the application of a small perturbing magnetic field $B_{RF}$ \cite{Bloom:62, Zigdon2010}. In this work we focus on the development of double-resonance magnetic sensors in which a single continuous-wave laser and magnetic perturbation coil are used, in order to design sensors that can be realised in small packages of low cost and power consumption.

Precession in the magnetisation of an atomic vapour can result in oscillating linear dichroism, detected by measurement of transmitted light using a polarimeter. Demodulation of this signal at the perturbation frequency $\omega_{RF}$ allows the measurement of a Lorentzian lineshape centred on the Larmor frequency $\omega_L$. This signal can be used to lock $\omega_{RF}$ to $\omega_L$ and thus the magnitude of $B_0$ can be tracked in real-time \cite{Bloom:62}. Alkali vapour ground state Larmor frequencies are in the range 150-400~kHz for geophysical fields, making double-resonance sensors attractive for measurements of the Earth's field. However, the amplitude of the resonant response is highly anisotropic with magnetic field orientation \cite{Pustelny2006,Ben-Kish2010} and, in the geophysical field range, heading errors (caused by shifts in the measured resonance centre due to unresolved nonlinear Zeeman spitting) can be significant \cite{Chalupczak2010,Alexandrov2003}.

The angular dependence of these potential systematic effects makes magnetometer signal characterisation under varying $B_0$ orientation an important step in development of unshielded sensors for practical applications. Several schemes for vector measurement of $\vec{B_0}$ have been developed \cite{Seltzer2004,Patton2014,Afach:15}. 
In this work we take advantage of the precise control of $B_0$ demonstrated in \cite{Ingleby2017} to measure the dependence of the double-resonance magnetometer signal amplitude and phase on the orientation of $B_0$ relative to the light propagation axis and $B_{RF}$. Our work focusses on single-wavelength small-modulation ($B_{RF} \ll B_0$) double-resonance magnetometry, as sensors designed with this technique can be realised using compact, portable hardware and inexpensive, scalable electronics. 

Precise measurements of $B_0$ orientation effects in transmission-based single-wavelength double-resonance magnetometry have been recently reported in \cite{Colombo2017}. By contrast with the transmission measurements reported in that publication, we report results obtained using a polarimeter to measure relative transmission of orthogonal linear polarisations, modifying the relation between the observed signal and the evolving atomic polarisation.

\section{Magnetic resonance in arbitrarily oriented fields}
\begin{figure}
\includegraphics[width=\linewidth]{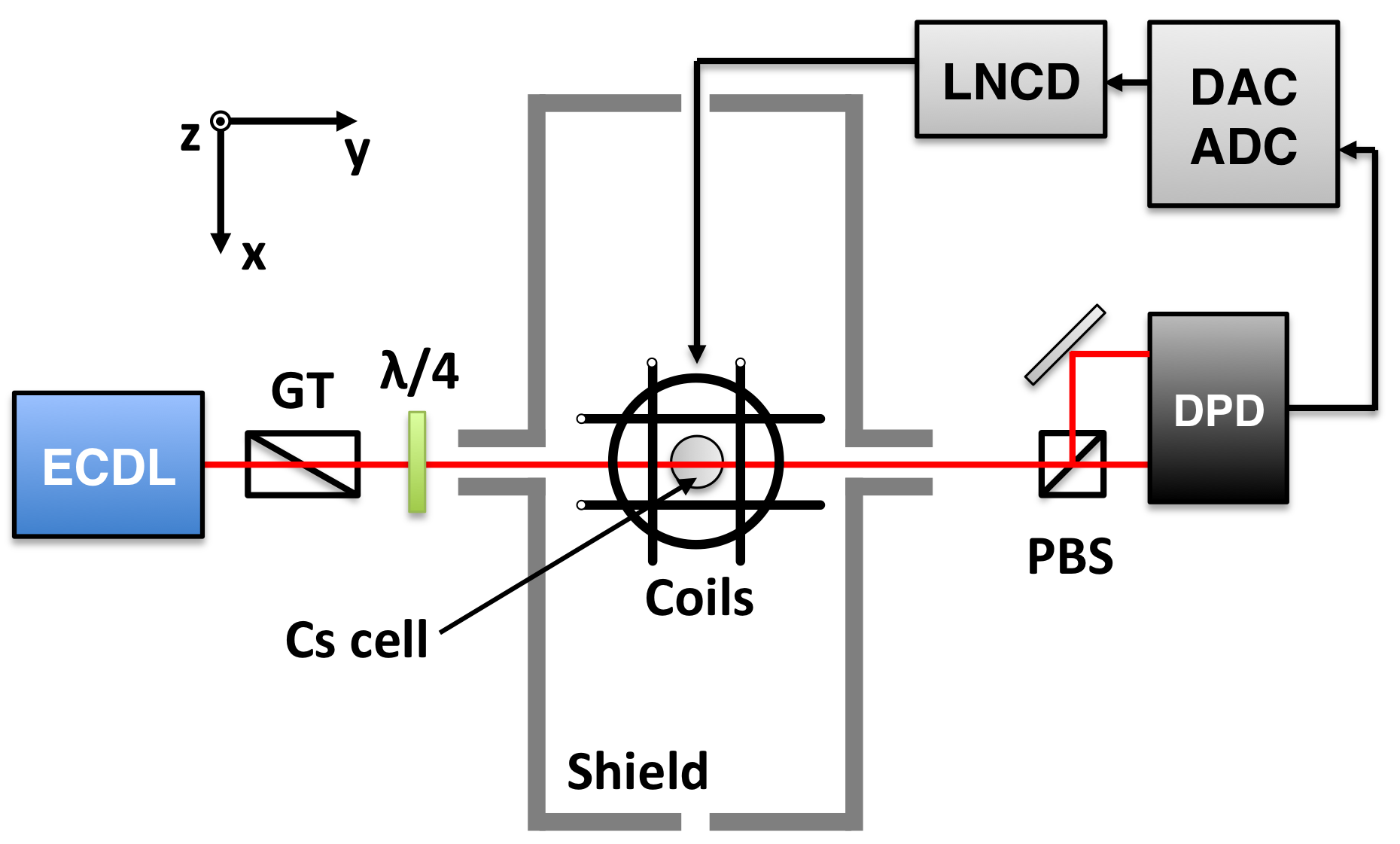}
\caption{Schematic of the apparatus used in this work, showing external cavity diode laser (ECDL), Glan-Thompson linear polariser (GT), quarter-wave plate ($\lambda/4$), magnetometer cell, five-layer mu-metal shield, three-axis Helmholtz coils, polarising beam splitter (PBS), differential photodetector (DPD), low-noise coil driver (LNCD) and data acquisition system (DAC/ADC). The data acquisition system is controlled using a PC (not shown).} 
\label{setup}
\end{figure}
Figure \ref{setup} shows the apparatus used for magnetic resonance measurements in a low-pressure antirelaxation-coated caesium vapour cell. The dimensions, fabrication details and characterisation of these cells are described in \cite{Castagna2009}. The magnitude and orientation of the static field $B_0$ can be controlled in software, with measured tolerances in field magnitude and orientation of 0.94~nT and 5.9~mrad, respectively. For a detailed description and evaluation of the static field control, see \cite{Ingleby2017}.

The $^{133}$Cs vapour is optically pumped using circularly polarised light resonant with the $6^2$S$_{1/2}$~$(F=4)$ to $6^2$P$_{1/2}$~$(F=3)$ transition, creating a net magnetisation in the Cs vapour. The beam has a circular profile of diameter 2~mm and total power 10~$\mu$W. Transitions between the Zeeman states of the Cs vapour $6^2$S$_{1/2}$~$(F=4)$ state are probed by the application of $B_{RF}$ using single-turn coils co-wound on the $y$ and $z$ Helmholtz coil formers. The diameters of these coils are 84~mm ($y$-axis) and 70~mm ($z$-axis). The amplitude of $B_{RF}$ is kept constant at 3~nT throughout the measurements. A polarimeter, comprising a polarising beam splitter and differential photodetector is used to detect changes in Cs magnetisation.

The magnetisation created in the Cs vapour by optical pumping is well described by the representation of the ground-state Zeeman level density operator $\rho$ using multipole moments $m_{k,q}$. We follow the detailed formalism of \cite{Weis2006}, writing 
\begin{equation}
\rho=\sum_{k=0}^{2F}\sum_{q=-k}^{k}m_{k,q}T_q^{(k)},
\end{equation}
where $T_q^{(k)}$ are irreducible basis tensors in the chosen quantisation basis. We consider a three-stage process to derive the observed polarimeter signal; the creation of atomic magnetisation through optical pumping, the evolution of that magnetisation in the magnetic field, and the observable effects on transmitted light. This three-stage model is only valid in the case of weak optical pumping (where the optical pumping rate is much smaller than the magnetic spin relaxation rate), in which the evolution of the magnetisation (stage two) can be considered due to the action of magnetic fields only. 

\begin{figure}
\includegraphics[width=\linewidth]{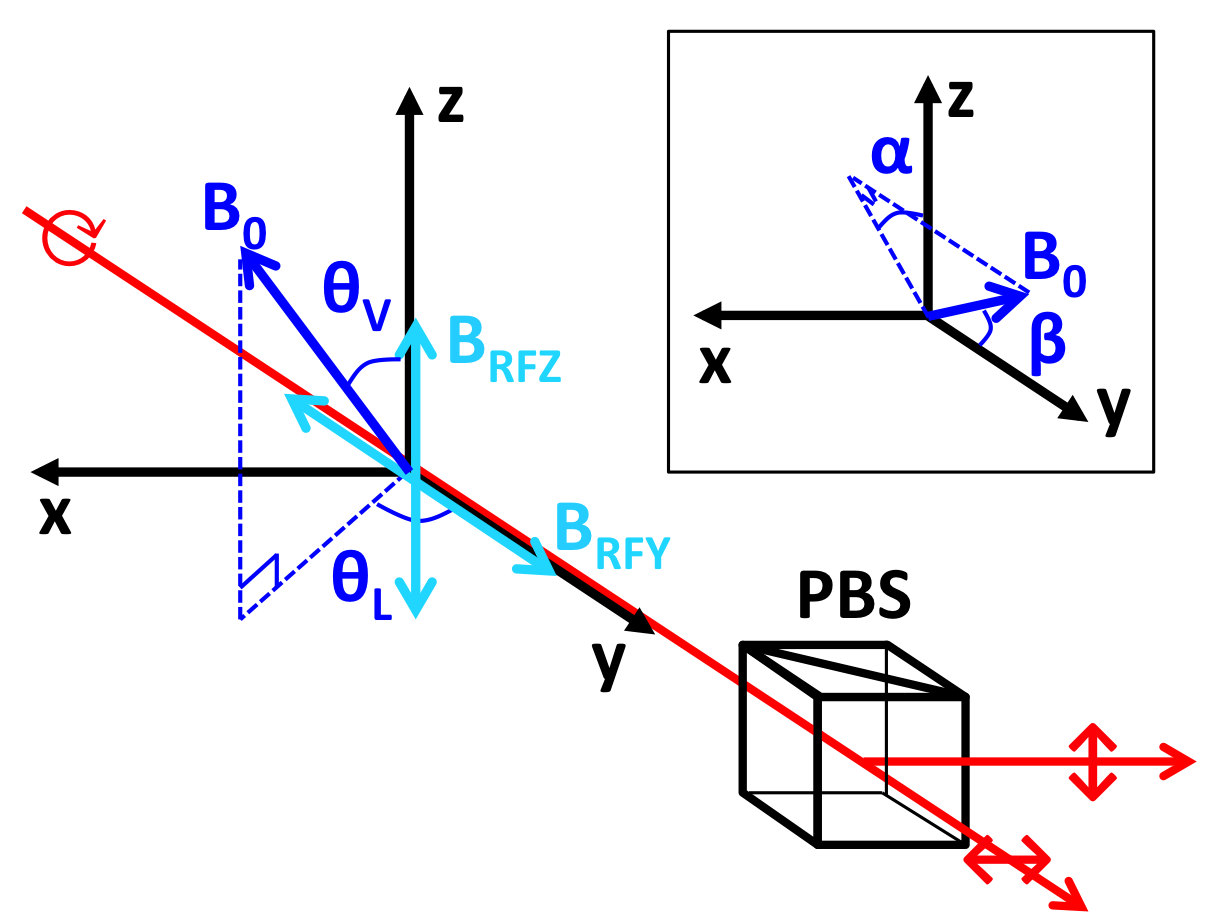}
\caption{Schematic showing the geometry of the magnetometer system in the laboratory frame. The orientation of the static field $B_0$ is defined by the angles to the light propagation and vertical axes, $\theta_L$ and $\theta_V$, respectively. Inset, alternative polar basis angles $\alpha$ and $\beta$ are defined. The orientations of the oscillating fields $B_{RFZ}$ and $B_{RFY}$, and the polarisation analyser (PBS), are also shown.}
\label{geometry}
\end{figure}

Figure \ref{geometry} shows the coordinate basis used to describe the laboratory frame. If the quantisation axis is parallel to the \textit{k}-vector of the pump light, then non-absorbing atomic dark states are described by sums of atomic multipoles in which only $m_{k,0}$ moments are non-zero, where $k$ may be any positive integer for circularly-polarised light and $k$ any even positive integer for linearly-polarised light \cite{Bevilacqua2014}. Optical pumping dynamically couples multipole moments of any $k,q$ to moments for which $q=0$ in this basis. In a static magnetic field the resulting equilibrium moments $m^{eq}_{k,q}$ depend on the rate of optical pumping and the orientation of the static field $B_0$ \cite{Bevilacqua2014}. 

Our magnetometer signal is the oscillating magnetisation response to $B_{RF}(\omega_{RF})$, so we define a rotating wave frame $x'-y'-z'$ (RW) such that $B_0$ lies along $z'$ and the component of $B_{RF}$ perpendicular to $B_0$ at $t=0$, $B_{RF}^\bot$, is in the negative $x'$ direction. In the presence of weak optical pumping, the atomic polarisation relaxes to states which are symmetric around the quantisation axis $B_0$, in which only $m'^{eq}_{k,0}$ are non-zero.

The evolution of $m_{k,q}$ in the presence of a magnetic field was derived from the Liouville equation in \cite{Weis2006}, and is given by
\begin{equation}
\frac{d}{dt}m_{k,q}=\sum_{q'}\mathbb{O}^{(k)}_{qq'} m_{k,q'} - \Gamma^{(k)}_{qq'} (m_{k,q'} - m^{eq}_{k,q'}),
\end{equation}
where $\mathbb{O}^{(k)}_{qq'}$ is a $(2k+1)\times (2k+1)$ matrix describing the action of $B_0$ and $B_{RF}$ on $m_{k,q}$ and $\Gamma^{(k)}_{qq'}$ is a $(2k+1)\times (2k+1)$ matrix describing the relaxation of $m_{k,q}$ to $m^{eq}_{k,q}$. By solving for equilibrium multipole moments in the RW frame $m'_{k,q}$, and transforming to the laboratory frame, we obtain expressions for the oscillating components of $m_{k,q}(t)$. Transformations of $m_{k,q}(t)$ under rotation are obtained using Wigner D-matrices, see \cite{Morrison1987} for details.

A balanced polarimeter is used for these measurements, ensuring that the observed signal is sensitive to relative changes in the transmission of orthogonal linearly polarised light through the cell. Signal contributions due to oscillating transmission of circularly polarised light will be observed in phase on both polarimeter channels, and will not contribute to the measured differential signal. The measured signal is thus modelled by the difference in absorption of the orthogonal linear polarisation states of the light received at the polarisation analyser (see Figure \ref{geometry}). The absorption coefficient $\kappa$ for linearly polarised light in a medium described by multipole moments $m_{k,q}(t)$ is proportional to 
\begin{equation}
\kappa \propto \frac{A_0}{\sqrt{3}}m_{0,0}-\sqrt{\frac{2}{3}}A_2 m_{2,0},
\end{equation}
where $A_0$ and $A_2$ are analysing powers \cite{Weis2006}. The analysing powers $A_0$, $A_2$ and monopole moment $m_{0,0}$ are invariant under rotations of basis, and cancel in subtraction, so we can write a function $f(t)$ which is proportional to the polarimeter signal,
\begin{equation}
\begin{aligned}
f(t) &= m_{2,0}(t)-m''_{2,0}(t) \\
&= \tfrac{3}{2}m_{2,0}(t)-\sqrt{\tfrac{3}{8}}(m_{2,2}(t) + m_{2,-2}(t)),
\end{aligned}
\end{equation}
where $m_{k,q}(t)$ denote multipole moments in the lab frame and $m''_{k,q}(t)$ multipole moments in a frame rotated such that the quantisation axis is coincident with the orthogonal linear polarisation axis ($x$-axis). Since the orthogonal linear states separated by the analyser are not sensitive to $m_{1,q}(t)$, and $m_{0,0}$ is constant in time, $k=2$ for the lowest-order multipole moments contributing to $f(t)$.

Following \cite{Weis2006}, and defining $x\equiv (\gamma B_0 - \omega_{RF})/\Gamma$ and $S\equiv \gamma B_{RF}^{\bot}/\Gamma$ for convenience, we obtain the following equations-of-motion for the RW-frame $k=2$ multipole moments,
\begin{equation}
\frac{i}{\Gamma}\dot{m}'_{2,q} = M_{qq'}m'_{2,q'}+i m'^{eq}_{2,q},
\end{equation}
where $\gamma$ is the gyromagnetic ratio for the Cs $6^2$S$_{1/2}$~$(F=4)$ ground state and 
\begin{equation}
M_{qq'} = \begin{pmatrix}2x-i&S&0&0&0\\
S&x-i&\sqrt{\tfrac{3}{2}}S&0&0\\
0&\sqrt{\tfrac{3}{2}}S&-i&\sqrt{\tfrac{3}{2}}S&0\\
0&0&\sqrt{\tfrac{3}{2}}S&-x-i&S\\
0&0&0&S&-2x-i\end{pmatrix}.
\end{equation}
We have assumed that relaxation in the system is isotropic with rate $\Gamma$, and that $B_{RF} \ll B_0$. 

Steady-state solutions for $m'_{2,q}$ in the RW frame are obtained by setting $\dot{m}'_{2,q} = 0$. After transformation to the laboratory frame we obtain terms in $e^{0 i\omega_{RF} t}$, $e^{1 i\omega_{RF} t}$ and $e^{2 i\omega_{RF} t}$, corresponding to DC, first-harmonic and second-harmonic resonant magnetisation responses to the perturbing field $B_{RF}$. 

\section{Anisotropy in resonant polarimeter signal response}

To measure the first-harmonic response to $B_{RF}(\omega_{RF})$, the polarimeter signal is digitised and demodulated in software with a reference phase-locked to $B_{RF}(t)$. 

Collecting terms from $f(t)$ in $\cos(\omega_{RF} t)$ and $\sin(\omega_{RF} t)$ we derive expressions for the in-phase, $X$, and quadrature, $Y$, components of the demodulated signal. 
The on-resonance ($x=0$) amplitudes of $X$ and $Y$ are strongly dependent on the orientation of $B_0$ and $B_{RF}$ relative to the light propagation axis. For $B_{RF}$ aligned along the $y$ and $z$ axes (denoted RFy and RFz respectively),
\begin{equation}
X_{RFy} = -\frac{3m'^{eq}_{2,0}S}{4S^2+1}\sin2\alpha\sin\beta,
\end{equation}
\begin{equation}
Y_{RFy} = \frac{3m'^{eq}_{2,0}S}{2(4S^2+1)}\cos2\alpha\sin2\beta,
\end{equation}
\begin{equation}
X_{RFz} = -\frac{3m'^{eq}_{2,0}S}{2(4S^2+1)}\sin2\theta_L\sin\theta_V,
\end{equation}
\begin{equation}
Y_{RFz} = \frac{3m'^{eq}_{2,0}S}{2(4S^2+1)}(1+\cos^2\theta_L)\sin2\theta_V,
\end{equation}
where $\cos\beta = \sin\theta_V \cos\theta_L$ and $\tan\alpha = \tan\theta_V \sin\theta_L$ define an alternative polar basis in the laboratory frame.

In practical magnetometry measurements we are concerned with the amplitude $R$ and phase $\phi$ of the magnetometer signal response. Defining
\begin{equation}
R^2 \equiv X^2+Y^2,
\end{equation}
\begin{equation}
\tan\phi \equiv X/Y,
\end{equation}
we obtain on-resonance signal amplitudes
\begin{equation}
R^2_{RFy} = A^2 \sin^22\beta \left(\cos^2\beta \cos^22\alpha +\sin^22\alpha \right), \label{Eq_RY}
\end{equation}
\begin{equation}
R^2_{RFz} = A^2 \sin^4\theta_V (4\cos^2\theta_V (1+\cos^2\theta_L)^2 +\sin^22\theta_L ), \label{Eq_RZ}
\end{equation}
and phases
\begin{equation}
\begin{aligned}
\tan\phi_{RFy} &= \frac{-2\sin2\alpha\sin\beta}{\cos2\alpha\sin2\beta} \\
&= \frac{2\sin\theta_L\tan\theta_V}{(\sin^2\theta_L\tan^2\theta_V-1)\cos\theta_L\sin\theta_V}, 
\label{Eq_PY}
\end{aligned}
\end{equation}
\begin{equation}
\tan\phi_{RFz} = \frac{-\sin2\theta_L\sin\theta_V}{(1+\cos^2\theta_L)\sin2\theta_V}, \label{Eq_PZ}
\end{equation}
where $A=\tfrac{3m'^{eq}_{2,0}S}{4(4S^2+1)}$. 

\begin{figure}
\begin{tabular}{c}
\includegraphics[width=0.95\linewidth]{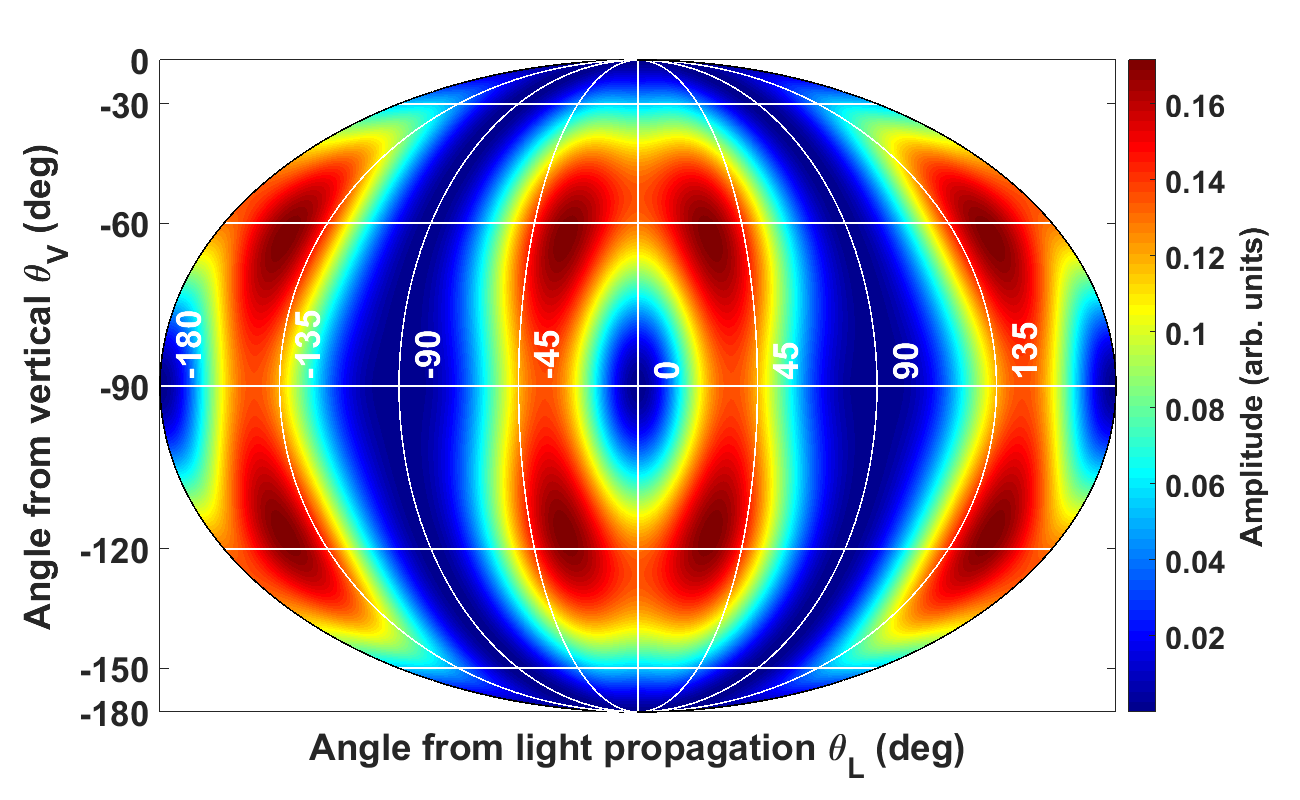} \\
\includegraphics[width=0.95\linewidth]{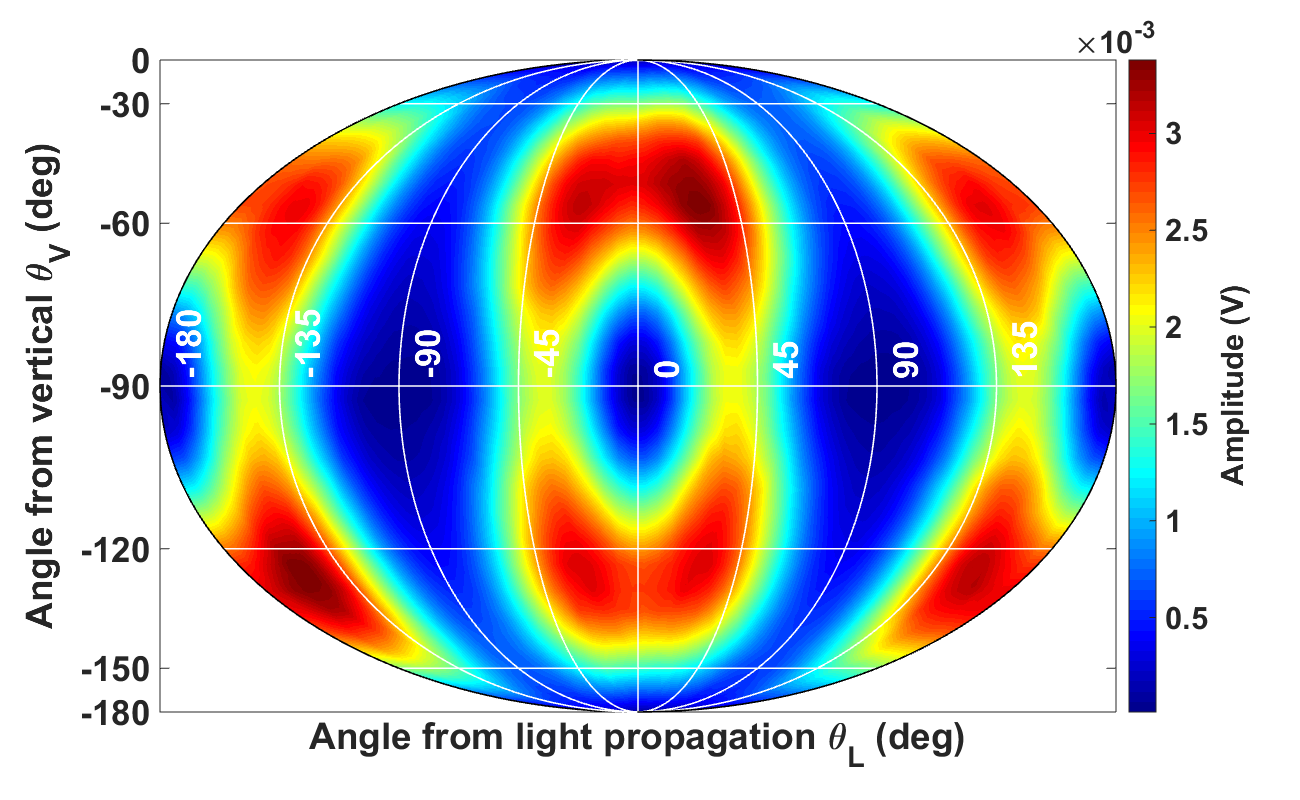}
\end{tabular}
\caption{Calculated (top) and observed (bottom) distribution of on-resonance first-harmonic polarimeter signal amplitude over $4\pi$ solid angle in response to a field modulation $B_{RF}$ applied parallel to the light propagation ($y$-)axis.}
\label{Fig_RY}
\end{figure}

\begin{figure}
\begin{tabular}{c}
\includegraphics[width=0.95\linewidth]{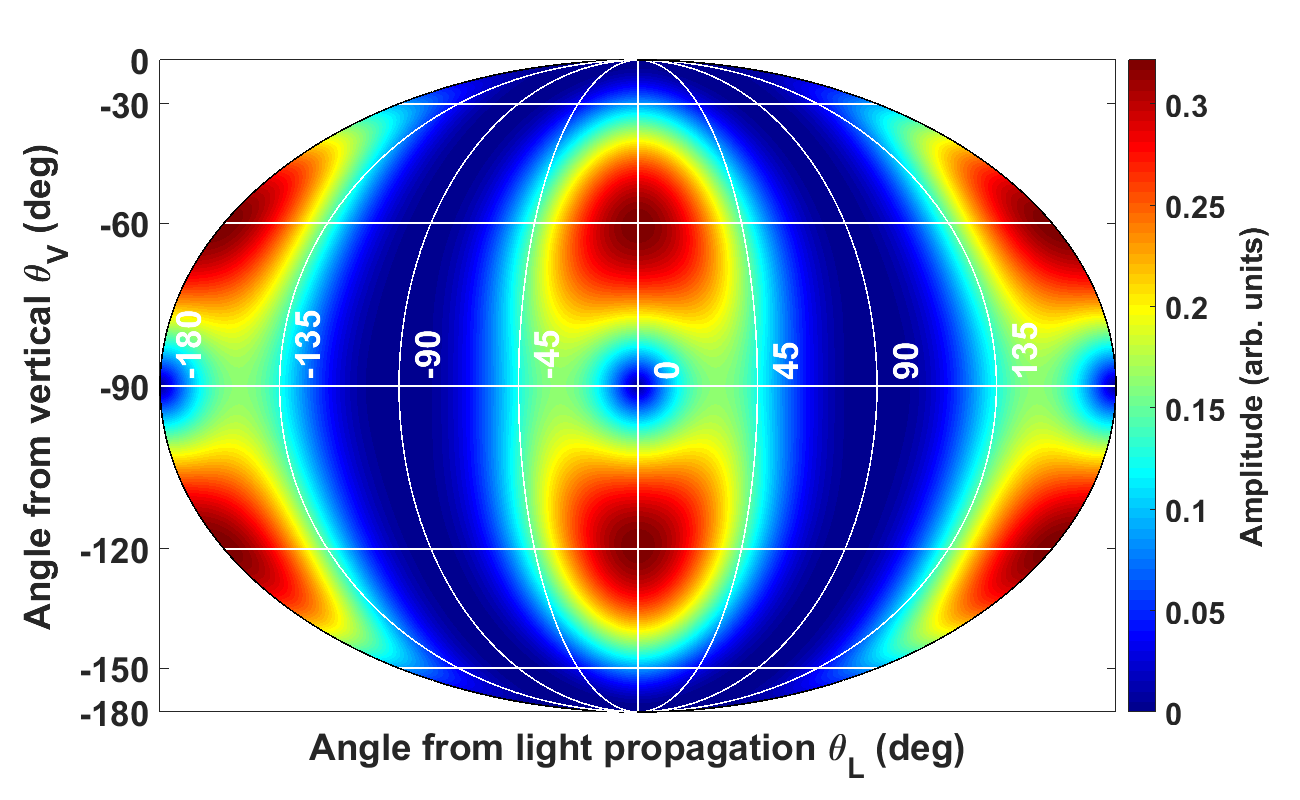} \\
\includegraphics[width=0.95\linewidth]{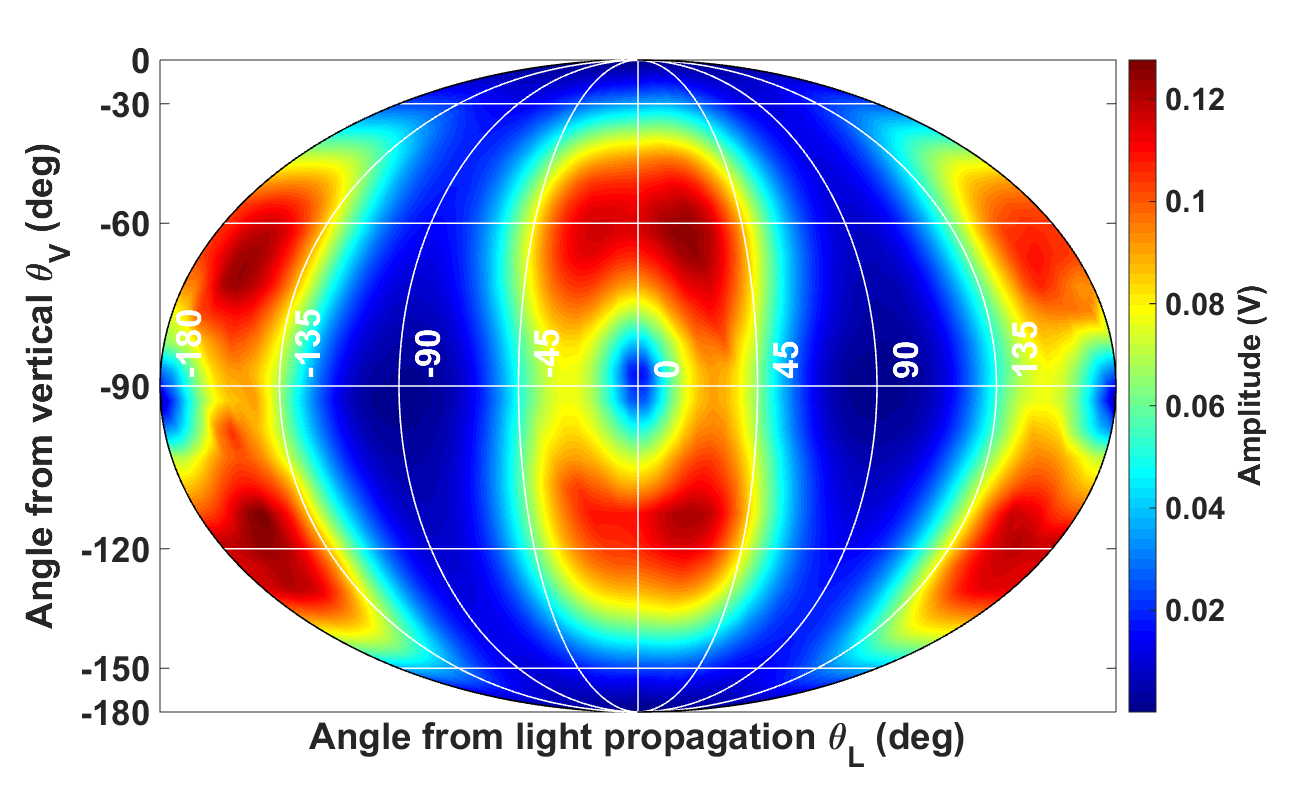}
\end{tabular}
\caption{Calculated (top) and observed (bottom) distribution of on-resonance first-harmonic polarimeter signal amplitude over $4\pi$ solid angle in response to a field modulation $B_{RF}$ applied parallel to the vertical ($z$-)axis.}
\label{Fig_RZ}
\end{figure}

\begin{figure}
\begin{tabular}{c}
\includegraphics[width=0.95\linewidth]{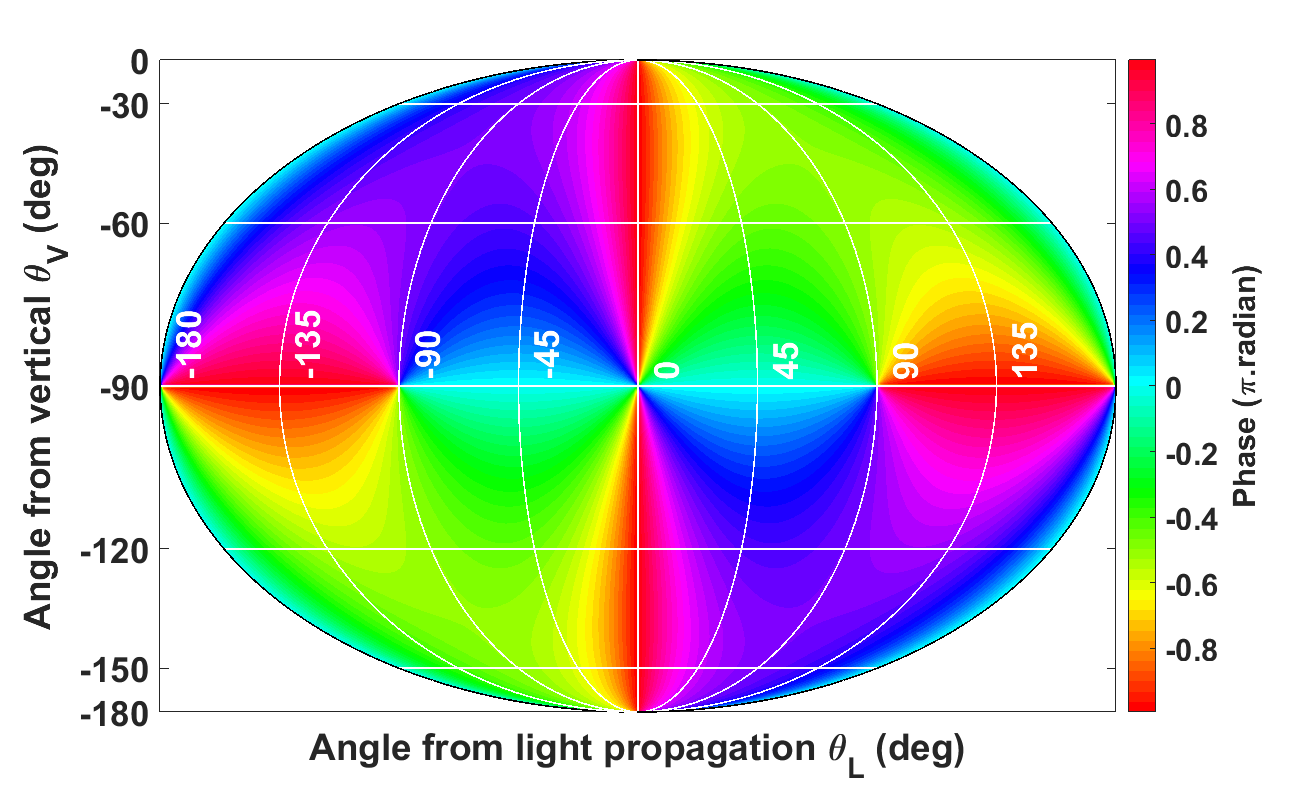} \\
\includegraphics[width=0.95\linewidth]{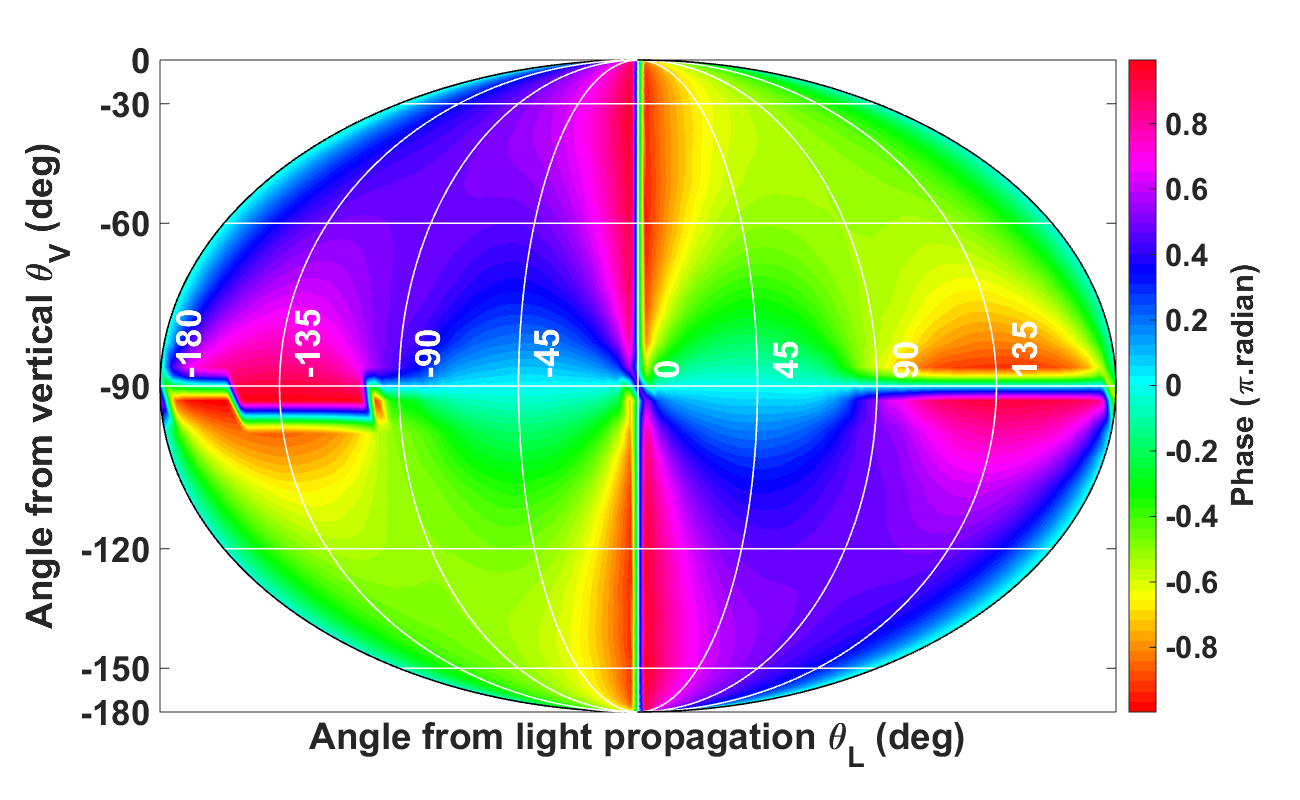}
\end{tabular}
\caption{Calculated (top) and observed (bottom) distribution of on-resonance first-harmonic polarimeter signal phase over $4\pi$ solid angle in response to a field modulation $B_{RF}$ applied parallel to the light propagation ($y$-)axis.}
\label{Fig_PY}
\end{figure}

\begin{figure}
\begin{tabular}{c}
\includegraphics[width=0.95\linewidth]{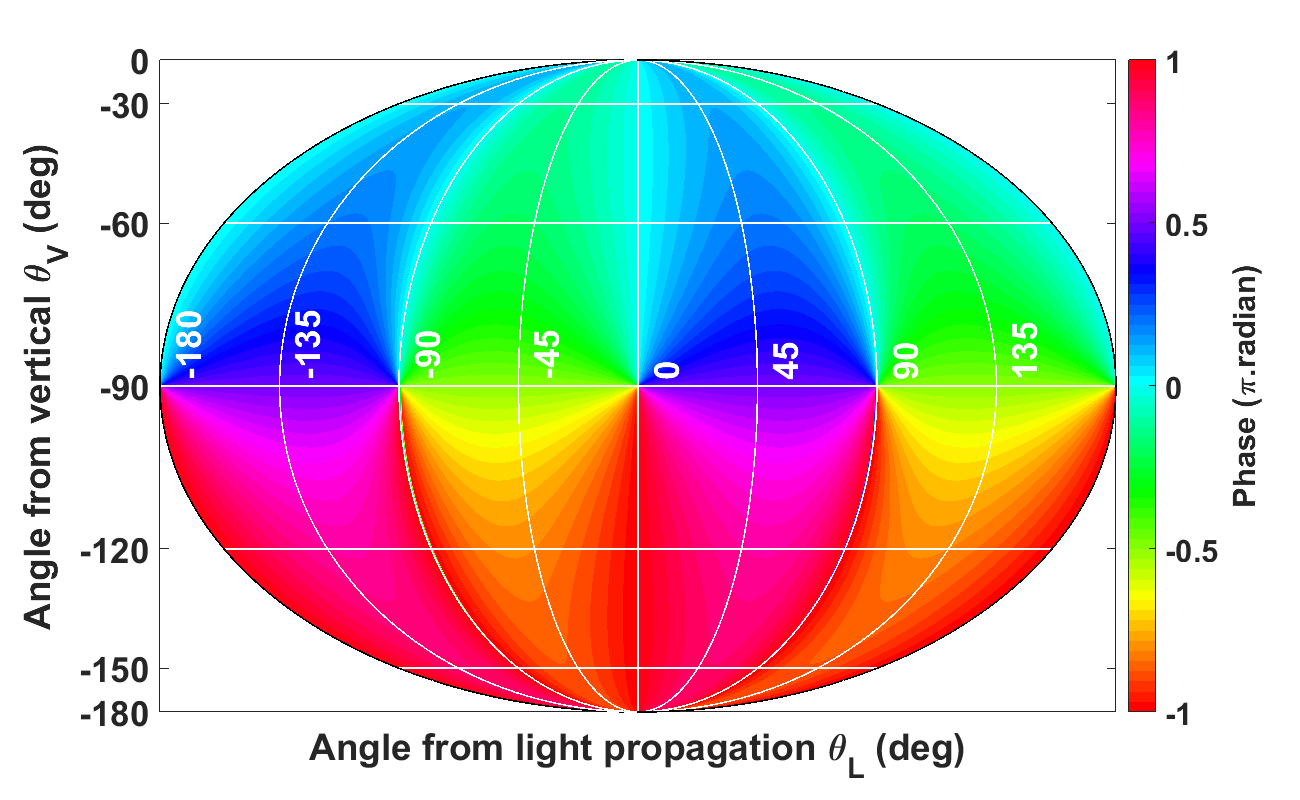} \\
\includegraphics[width=0.95\linewidth]{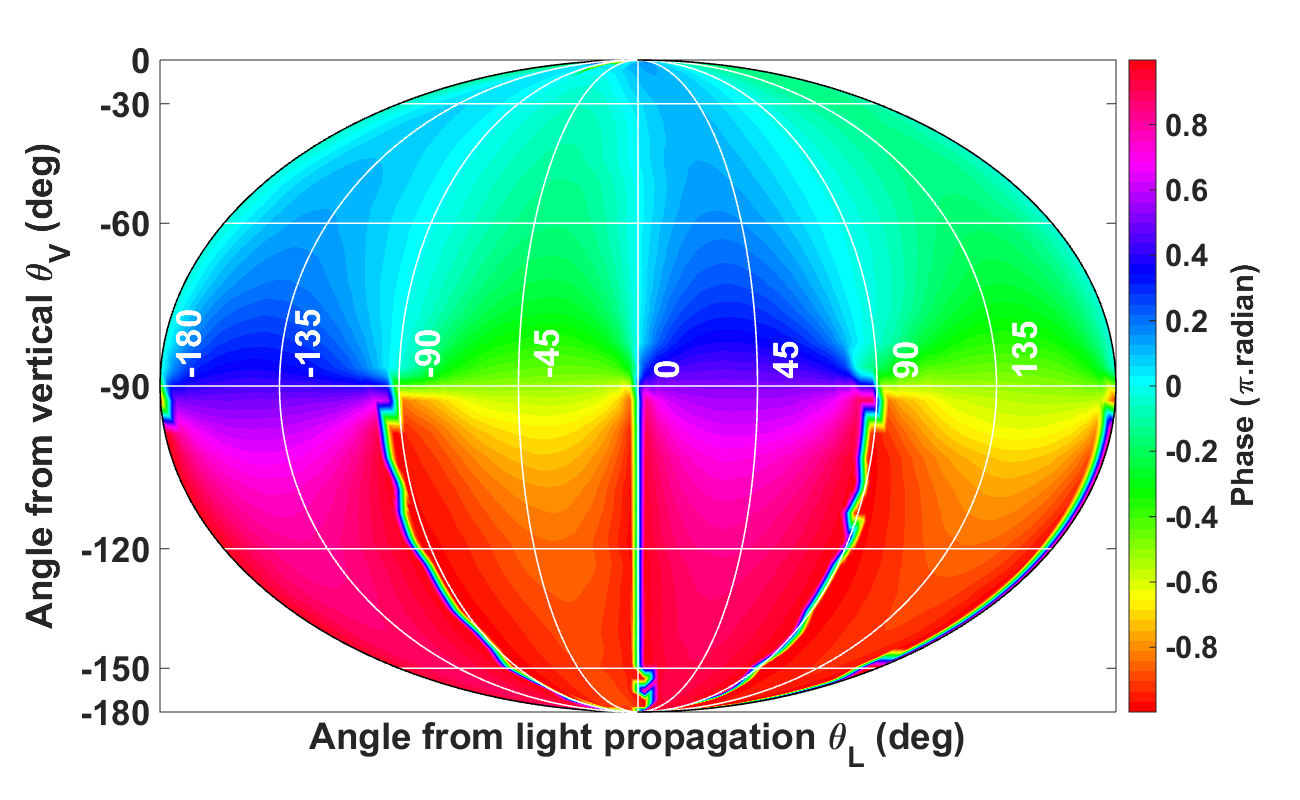}
\end{tabular}
\caption{Calculated (top) and observed (bottom) distribution of on-resonance first-harmonic polarimeter signal phase over $4\pi$ solid angle in response to a field modulation $B_{RF}$ applied parallel to the vertical ($z$-)axis.}
\label{Fig_PZ}
\end{figure}

Using the $B_0$ control, signal demodulation and resonance lineshape fitting described in \cite{Ingleby2017}, angular scans of the measured on-resonance signal amplitude $R$ and phase $\phi$ were made. Each angular scan comprised 1646 varying orientations of $B_0 =$ 200~nT spread with equal angular density over the full $4\pi$ solid angle. Figures \ref{Fig_RY}-\ref{Fig_PZ} show these data plotted alongside distributions calculated using Equations \ref{Eq_RY}-\ref{Eq_PZ}.

\section{Conclusions}

Equations \ref{Eq_RY}-\ref{Eq_RZ} model the observed dead-zones (orientations of $B_0$ for which $R\to0$) and symmetries in the distribution of observed signal amplitude $R$ (Figures \ref{Fig_RY}-\ref{Fig_RZ}). We note that our experimental measurements of signal amplitude differ slightly in shape to the predicted distributions, a feature which could be reduced by inclusion of higher-order multipole moments in the model signal calculation. In the case where the polarimeter is not perfectly balanced, the signal contribution due to the evolution of first-order multipole moments $m_{1,q}$, which modulate the transmission of circularly polarised light, will not be fully cancelled by the differential photodetector. This may also account for discrepancies between the observed and calculated distributions of $R$.

Equations \ref{Eq_PY}-\ref{Eq_PZ} appear to model the observed polarimeter signal phase well, and the observed data confirm the strong correlation of on-resonance phase with $B_0$ orientation (Figures \ref{Fig_PY}-\ref{Fig_PZ}). The clarity and good agreement of the measured data with calculated signal phase may be due in part to the cancellation of spurious signal terms in the phase calculation $\phi = \arctan (X/Y)$. We observe from Figures \ref{Fig_PY} and \ref{Fig_PZ} that $\phi_{RFy}$ and $\phi_{RFz}$ are uniquely mapped to $\theta_L$ and $\theta_V$, but the presence of asymptotes and stationary values in $\phi_{RFy}(\theta_L,\theta_V)$ and $\phi_{RFz}(\theta_L,\theta_V)$ will make calculation of the full $B_0$ vector using measured $\phi_{RFy}$ and $\phi_{RFz}$ non-trivial. However, it may be possible to exploit angular information in the observed signal phase for magnetometer sensitivity optimisation through active feedback to either the local field or mechanical orientation of the sensor.

\begin{acknowledgements}
\section{Acknowledgements}
The authors would like to thank Prof. Antoine Weis and Dr. Victor Lebedev of Fribourg University for supplying the Cs vapour cell used in this work, and would also like to thank Dr. Jonathan Pritchard for stimulating discussions. This work was funded by the UK Quantum Technology Hub in Sensing and Metrology, EPSRC (EP/M013294/1).

The modulation design and vector measurement scheme described in this paper are the subject of UK Patent Application No. 1706674.7 (University of Strathclyde).
\end{acknowledgements}

\bibliography{vector_phase_paper.bib}

\begin{thebibliography}{20}%
\makeatletter
\providecommand \@ifxundefined [1]{%
 \@ifx{#1\undefined}
}%
\providecommand \@ifnum [1]{%
 \ifnum #1\expandafter \@firstoftwo
 \else \expandafter \@secondoftwo
 \fi
}%
\providecommand \@ifx [1]{%
 \ifx #1\expandafter \@firstoftwo
 \else \expandafter \@secondoftwo
 \fi
}%
\providecommand \natexlab [1]{#1}%
\providecommand \enquote  [1]{``#1''}%
\providecommand \bibnamefont  [1]{#1}%
\providecommand \bibfnamefont [1]{#1}%
\providecommand \citenamefont [1]{#1}%
\providecommand \href@noop [0]{\@secondoftwo}%
\providecommand \href [0]{\begingroup \@sanitize@url \@href}%
\providecommand \@href[1]{\@@startlink{#1}\@@href}%
\providecommand \@@href[1]{\endgroup#1\@@endlink}%
\providecommand \@sanitize@url [0]{\catcode `\\12\catcode `\$12\catcode
  `\&12\catcode `\#12\catcode `\^12\catcode `\_12\catcode `\%12\relax}%
\providecommand \@@startlink[1]{}%
\providecommand \@@endlink[0]{}%
\providecommand \url  [0]{\begingroup\@sanitize@url \@url }%
\providecommand \@url [1]{\endgroup\@href {#1}{\urlprefix }}%
\providecommand \urlprefix  [0]{URL }%
\providecommand \Eprint [0]{\href }%
\providecommand \doibase [0]{http://dx.doi.org/}%
\providecommand \selectlanguage [0]{\@gobble}%
\providecommand \bibinfo  [0]{\@secondoftwo}%
\providecommand \bibfield  [0]{\@secondoftwo}%
\providecommand \translation [1]{[#1]}%
\providecommand \BibitemOpen [0]{}%
\providecommand \bibitemStop [0]{}%
\providecommand \bibitemNoStop [0]{.\EOS\space}%
\providecommand \EOS [0]{\spacefactor3000\relax}%
\providecommand \BibitemShut  [1]{\csname bibitem#1\endcsname}%
\let\auto@bib@innerbib\@empty
\bibitem [{\citenamefont {Sheng}\ \emph {et~al.}(2013)\citenamefont {Sheng},
  \citenamefont {Li}, \citenamefont {Dural},\ and\ \citenamefont
  {Romalis}}]{Sheng2013}%
  \BibitemOpen
  \bibfield  {author} {\bibinfo {author} {\bibfnamefont {D.}~\bibnamefont
  {Sheng}}, \bibinfo {author} {\bibfnamefont {S.}~\bibnamefont {Li}}, \bibinfo
  {author} {\bibfnamefont {N.}~\bibnamefont {Dural}}, \ and\ \bibinfo {author}
  {\bibfnamefont {M.~V.}\ \bibnamefont {Romalis}},\ }\href {\doibase
  10.1103/PhysRevLett.110.160802} {\bibfield  {journal} {\bibinfo  {journal}
  {Phys. Rev. Lett.}\ }\textbf {\bibinfo {volume} {110}},\ \bibinfo {pages}
  {160802} (\bibinfo {year} {2013})},\ \Eprint {http://arxiv.org/abs/1208.1099}
  {arXiv:1208.1099} \BibitemShut {NoStop}%
\bibitem [{\citenamefont {Schwindt}\ \emph {et~al.}(2007)\citenamefont
  {Schwindt}, \citenamefont {Lindseth}, \citenamefont {Knappe}, \citenamefont
  {Shah}, \citenamefont {Kitching},\ and\ \citenamefont {Liew}}]{Schwindt2007}%
  \BibitemOpen
  \bibfield  {author} {\bibinfo {author} {\bibfnamefont {P.~D.~D.}\
  \bibnamefont {Schwindt}}, \bibinfo {author} {\bibfnamefont {B.}~\bibnamefont
  {Lindseth}}, \bibinfo {author} {\bibfnamefont {S.}~\bibnamefont {Knappe}},
  \bibinfo {author} {\bibfnamefont {V.}~\bibnamefont {Shah}}, \bibinfo {author}
  {\bibfnamefont {J.}~\bibnamefont {Kitching}}, \ and\ \bibinfo {author}
  {\bibfnamefont {L.-A.}\ \bibnamefont {Liew}},\ }\href {\doibase
  10.1063/1.2709532} {\bibfield  {journal} {\bibinfo  {journal} {Appl. Phys.
  Lett.}\ }\textbf {\bibinfo {volume} {90}},\ \bibinfo {pages} {081102}
  (\bibinfo {year} {2007})}\BibitemShut {NoStop}%
\bibitem [{\citenamefont {Bell}\ and\ \citenamefont {Bloom}(1961)}]{Bell1961}%
  \BibitemOpen
  \bibfield  {author} {\bibinfo {author} {\bibfnamefont {W.~E.}\ \bibnamefont
  {Bell}}\ and\ \bibinfo {author} {\bibfnamefont {A.~L.}\ \bibnamefont
  {Bloom}},\ }\href {\doibase 10.1103/PhysRevLett.6.280} {\bibfield  {journal}
  {\bibinfo  {journal} {Phys. Rev. Lett.}\ }\textbf {\bibinfo {volume} {6}},\
  \bibinfo {pages} {280} (\bibinfo {year} {1961})}\BibitemShut {NoStop}%
\bibitem [{\citenamefont {Jim{\'{e}}nez-Mart{\'{i}}nez}\ \emph
  {et~al.}(2010)\citenamefont {Jim{\'{e}}nez-Mart{\'{i}}nez}, \citenamefont
  {Griffith}, \citenamefont {Wang}, \citenamefont {Knappe}, \citenamefont
  {Kitching}, \citenamefont {Smith},\ and\ \citenamefont
  {Prouty}}]{Jimenez-Martinez2010}%
  \BibitemOpen
  \bibfield  {author} {\bibinfo {author} {\bibfnamefont {R.}~\bibnamefont
  {Jim{\'{e}}nez-Mart{\'{i}}nez}}, \bibinfo {author} {\bibfnamefont {W.~C.}\
  \bibnamefont {Griffith}}, \bibinfo {author} {\bibfnamefont {Y.~J.}\
  \bibnamefont {Wang}}, \bibinfo {author} {\bibfnamefont {S.}~\bibnamefont
  {Knappe}}, \bibinfo {author} {\bibfnamefont {J.}~\bibnamefont {Kitching}},
  \bibinfo {author} {\bibfnamefont {K.}~\bibnamefont {Smith}}, \ and\ \bibinfo
  {author} {\bibfnamefont {M.~D.}\ \bibnamefont {Prouty}},\ }\href {\doibase
  10.1109/TIM.2009.2023829} {\bibfield  {journal} {\bibinfo  {journal} {IEEE
  Trans. Instrum. Meas.}\ }\textbf {\bibinfo {volume} {59}},\ \bibinfo {pages}
  {372} (\bibinfo {year} {2010})}\BibitemShut {NoStop}%
\bibitem [{\citenamefont {Breschi}\ \emph {et~al.}(2014)\citenamefont
  {Breschi}, \citenamefont {Gruji{\'{c}}}, \citenamefont {Knowles},\ and\
  \citenamefont {Weis}}]{Breschi2014}%
  \BibitemOpen
  \bibfield  {author} {\bibinfo {author} {\bibfnamefont {E.}~\bibnamefont
  {Breschi}}, \bibinfo {author} {\bibfnamefont {Z.~D.}\ \bibnamefont
  {Gruji{\'{c}}}}, \bibinfo {author} {\bibfnamefont {P.}~\bibnamefont
  {Knowles}}, \ and\ \bibinfo {author} {\bibfnamefont {A.}~\bibnamefont
  {Weis}},\ }\href {\doibase 10.1063/1.4861458} {\bibfield  {journal} {\bibinfo
   {journal} {Appl. Phys. Lett.}\ }\textbf {\bibinfo {volume} {104}} (\bibinfo
  {year} {2014}),\ 10.1063/1.4861458},\ \Eprint
  {http://arxiv.org/abs/arXiv:1312.3567v1} {arXiv:arXiv:1312.3567v1}
  \BibitemShut {NoStop}%
\bibitem [{\citenamefont {Bloom}(1962)}]{Bloom:62}%
  \BibitemOpen
  \bibfield  {author} {\bibinfo {author} {\bibfnamefont {A.~L.}\ \bibnamefont
  {Bloom}},\ }\href {\doibase 10.1364/AO.1.000061} {\bibfield  {journal}
  {\bibinfo  {journal} {Appl. Opt.}\ }\textbf {\bibinfo {volume} {1}},\
  \bibinfo {pages} {61} (\bibinfo {year} {1962})}\BibitemShut {NoStop}%
\bibitem [{\citenamefont {Zigdon}\ \emph {et~al.}(2010)\citenamefont {Zigdon},
  \citenamefont {Wilson-Gordon}, \citenamefont {Guttikonda}, \citenamefont
  {Bahr}, \citenamefont {Neitzke}, \citenamefont {Rochester},\ and\
  \citenamefont {Budker}}]{Zigdon2010}%
  \BibitemOpen
  \bibfield  {author} {\bibinfo {author} {\bibfnamefont {T.}~\bibnamefont
  {Zigdon}}, \bibinfo {author} {\bibfnamefont {A.~D.}\ \bibnamefont
  {Wilson-Gordon}}, \bibinfo {author} {\bibfnamefont {S.}~\bibnamefont
  {Guttikonda}}, \bibinfo {author} {\bibfnamefont {E.~J.}\ \bibnamefont
  {Bahr}}, \bibinfo {author} {\bibfnamefont {O.}~\bibnamefont {Neitzke}},
  \bibinfo {author} {\bibfnamefont {S.~M.}\ \bibnamefont {Rochester}}, \ and\
  \bibinfo {author} {\bibfnamefont {D.}~\bibnamefont {Budker}},\ }\href
  {\doibase 10.1364/OE.18.025494} {\bibfield  {journal} {\bibinfo  {journal}
  {Opt. Express}\ }\textbf {\bibinfo {volume} {18}},\ \bibinfo {pages} {25494}
  (\bibinfo {year} {2010})},\ \Eprint {http://arxiv.org/abs/1008.3000}
  {arXiv:1008.3000} \BibitemShut {NoStop}%
\bibitem [{\citenamefont {Pustelny}\ \emph {et~al.}(2006)\citenamefont
  {Pustelny}, \citenamefont {Gawlik}, \citenamefont {Rochester}, \citenamefont
  {Kimball}, \citenamefont {Yashchuk},\ and\ \citenamefont
  {Budker}}]{Pustelny2006}%
  \BibitemOpen
  \bibfield  {author} {\bibinfo {author} {\bibfnamefont {S.}~\bibnamefont
  {Pustelny}}, \bibinfo {author} {\bibfnamefont {W.}~\bibnamefont {Gawlik}},
  \bibinfo {author} {\bibfnamefont {S.~M.}\ \bibnamefont {Rochester}}, \bibinfo
  {author} {\bibfnamefont {D.~F.~J.}\ \bibnamefont {Kimball}}, \bibinfo
  {author} {\bibfnamefont {V.~V.}\ \bibnamefont {Yashchuk}}, \ and\ \bibinfo
  {author} {\bibfnamefont {D.}~\bibnamefont {Budker}},\ }\href {\doibase
  10.1103/PhysRevA.74.063420} {\bibfield  {journal} {\bibinfo  {journal} {Phys.
  Rev. A}\ }\textbf {\bibinfo {volume} {74}},\ \bibinfo {pages} {063420}
  (\bibinfo {year} {2006})},\ \Eprint {http://arxiv.org/abs/0606257}
  {arXiv:0606257 [physics]} \BibitemShut {NoStop}%
\bibitem [{\citenamefont {Ben-Kish}\ and\ \citenamefont
  {Romalis}(2010)}]{Ben-Kish2010}%
  \BibitemOpen
  \bibfield  {author} {\bibinfo {author} {\bibfnamefont {A.}~\bibnamefont
  {Ben-Kish}}\ and\ \bibinfo {author} {\bibfnamefont {M.~V.}\ \bibnamefont
  {Romalis}},\ }\href {\doibase 10.1103/PhysRevLett.105.193601} {\bibfield
  {journal} {\bibinfo  {journal} {Phys. Rev. Lett.}\ }\textbf {\bibinfo
  {volume} {105}},\ \bibinfo {pages} {193601} (\bibinfo {year}
  {2010})}\BibitemShut {NoStop}%
\bibitem [{\citenamefont {Chalupczak}\ \emph {et~al.}(2010)\citenamefont
  {Chalupczak}, \citenamefont {Wojciechowski}, \citenamefont {Pustelny},\ and\
  \citenamefont {Gawlik}}]{Chalupczak2010}%
  \BibitemOpen
  \bibfield  {author} {\bibinfo {author} {\bibfnamefont {W.}~\bibnamefont
  {Chalupczak}}, \bibinfo {author} {\bibfnamefont {A.}~\bibnamefont
  {Wojciechowski}}, \bibinfo {author} {\bibfnamefont {S.}~\bibnamefont
  {Pustelny}}, \ and\ \bibinfo {author} {\bibfnamefont {W.}~\bibnamefont
  {Gawlik}},\ }\href {\doibase 10.1103/PhysRevA.82.023417} {\bibfield
  {journal} {\bibinfo  {journal} {Phys. Rev. A}\ }\textbf {\bibinfo {volume}
  {82}},\ \bibinfo {pages} {023417} (\bibinfo {year} {2010})}\BibitemShut
  {NoStop}%
\bibitem [{\citenamefont {Alexandrov}(2003)}]{Alexandrov2003}%
  \BibitemOpen
  \bibfield  {author} {\bibinfo {author} {\bibfnamefont {E.~B.}\ \bibnamefont
  {Alexandrov}},\ }\href {http://stacks.iop.org/1402-4896/2003/i=T105/a=005}
  {\bibfield  {journal} {\bibinfo  {journal} {Physica Scripta}\ }\textbf
  {\bibinfo {volume} {2003}},\ \bibinfo {pages} {27} (\bibinfo {year}
  {2003})}\BibitemShut {NoStop}%
\bibitem [{\citenamefont {Seltzer}\ and\ \citenamefont
  {Romalis}(2004)}]{Seltzer2004}%
  \BibitemOpen
  \bibfield  {author} {\bibinfo {author} {\bibfnamefont {S.~J.}\ \bibnamefont
  {Seltzer}}\ and\ \bibinfo {author} {\bibfnamefont {M.~V.}\ \bibnamefont
  {Romalis}},\ }\href {\doibase 10.1063/1.1814434} {\bibfield  {journal}
  {\bibinfo  {journal} {Appl. Phys. Lett.}\ }\textbf {\bibinfo {volume} {85}},\
  \bibinfo {pages} {4804} (\bibinfo {year} {2004})}\BibitemShut {NoStop}%
\bibitem [{\citenamefont {Patton}\ \emph {et~al.}(2014)\citenamefont {Patton},
  \citenamefont {Zhivun}, \citenamefont {Hovde},\ and\ \citenamefont
  {Budker}}]{Patton2014}%
  \BibitemOpen
  \bibfield  {author} {\bibinfo {author} {\bibfnamefont {B.}~\bibnamefont
  {Patton}}, \bibinfo {author} {\bibfnamefont {E.}~\bibnamefont {Zhivun}},
  \bibinfo {author} {\bibfnamefont {D.}~\bibnamefont {Hovde}}, \ and\ \bibinfo
  {author} {\bibfnamefont {D.}~\bibnamefont {Budker}},\ }\href {\doibase
  10.1103/PhysRevLett.113.013001} {\bibfield  {journal} {\bibinfo  {journal}
  {Phys. Rev. Lett.}\ }\textbf {\bibinfo {volume} {113}},\ \bibinfo {pages}
  {013001} (\bibinfo {year} {2014})}\BibitemShut {NoStop}%
\bibitem [{\citenamefont {Afach}\ \emph {et~al.}(2015)\citenamefont {Afach},
  \citenamefont {Ban}, \citenamefont {Bison}, \citenamefont {Bodek},
  \citenamefont {Chowdhuri}, \citenamefont {Gruji{\'{c}}}, \citenamefont
  {Hayen}, \citenamefont {H{\'{e}}laine}, \citenamefont {Kasprzak},
  \citenamefont {Kirch}, \citenamefont {Knowles}, \citenamefont {Koch},
  \citenamefont {Komposch}, \citenamefont {Kozela}, \citenamefont {Krempel},
  \citenamefont {Lauss}, \citenamefont {Lefort}, \citenamefont {Lemi{\`{e}}re},
  \citenamefont {Mtchedlishvili}, \citenamefont {Naviliat-Cuncic},
  \citenamefont {Piegsa}, \citenamefont {Prashanth}, \citenamefont
  {Qu{\'{e}}m{\'{e}}ner}, \citenamefont {Rawlik}, \citenamefont {Ries},
  \citenamefont {Roccia}, \citenamefont {Rozpedzik}, \citenamefont
  {Schmidt-Wellenburg}, \citenamefont {Severjins}, \citenamefont {Weis},
  \citenamefont {Wursten}, \citenamefont {Wyszynski}, \citenamefont {Zejma},\
  and\ \citenamefont {Zsigmond}}]{Afach:15}%
  \BibitemOpen
  \bibfield  {author} {\bibinfo {author} {\bibfnamefont {S.}~\bibnamefont
  {Afach}}, \bibinfo {author} {\bibfnamefont {G.}~\bibnamefont {Ban}}, \bibinfo
  {author} {\bibfnamefont {G.}~\bibnamefont {Bison}}, \bibinfo {author}
  {\bibfnamefont {K.}~\bibnamefont {Bodek}}, \bibinfo {author} {\bibfnamefont
  {Z.}~\bibnamefont {Chowdhuri}}, \bibinfo {author} {\bibfnamefont {Z.~D.}\
  \bibnamefont {Gruji{\'{c}}}}, \bibinfo {author} {\bibfnamefont
  {L.}~\bibnamefont {Hayen}}, \bibinfo {author} {\bibfnamefont
  {V.}~\bibnamefont {H{\'{e}}laine}}, \bibinfo {author} {\bibfnamefont
  {M.}~\bibnamefont {Kasprzak}}, \bibinfo {author} {\bibfnamefont
  {K.}~\bibnamefont {Kirch}}, \bibinfo {author} {\bibfnamefont
  {P.}~\bibnamefont {Knowles}}, \bibinfo {author} {\bibfnamefont {H.-C.}\
  \bibnamefont {Koch}}, \bibinfo {author} {\bibfnamefont {S.}~\bibnamefont
  {Komposch}}, \bibinfo {author} {\bibfnamefont {A.}~\bibnamefont {Kozela}},
  \bibinfo {author} {\bibfnamefont {J.}~\bibnamefont {Krempel}}, \bibinfo
  {author} {\bibfnamefont {B.}~\bibnamefont {Lauss}}, \bibinfo {author}
  {\bibfnamefont {T.}~\bibnamefont {Lefort}}, \bibinfo {author} {\bibfnamefont
  {Y.}~\bibnamefont {Lemi{\`{e}}re}}, \bibinfo {author} {\bibfnamefont
  {A.}~\bibnamefont {Mtchedlishvili}}, \bibinfo {author} {\bibfnamefont
  {O.}~\bibnamefont {Naviliat-Cuncic}}, \bibinfo {author} {\bibfnamefont
  {F.~M.}\ \bibnamefont {Piegsa}}, \bibinfo {author} {\bibfnamefont {P.~N.}\
  \bibnamefont {Prashanth}}, \bibinfo {author} {\bibfnamefont {G.}~\bibnamefont
  {Qu{\'{e}}m{\'{e}}ner}}, \bibinfo {author} {\bibfnamefont {M.}~\bibnamefont
  {Rawlik}}, \bibinfo {author} {\bibfnamefont {D.}~\bibnamefont {Ries}},
  \bibinfo {author} {\bibfnamefont {S.}~\bibnamefont {Roccia}}, \bibinfo
  {author} {\bibfnamefont {D.}~\bibnamefont {Rozpedzik}}, \bibinfo {author}
  {\bibfnamefont {P.}~\bibnamefont {Schmidt-Wellenburg}}, \bibinfo {author}
  {\bibfnamefont {N.}~\bibnamefont {Severjins}}, \bibinfo {author}
  {\bibfnamefont {A.}~\bibnamefont {Weis}}, \bibinfo {author} {\bibfnamefont
  {E.}~\bibnamefont {Wursten}}, \bibinfo {author} {\bibfnamefont
  {G.}~\bibnamefont {Wyszynski}}, \bibinfo {author} {\bibfnamefont
  {J.}~\bibnamefont {Zejma}}, \ and\ \bibinfo {author} {\bibfnamefont
  {G.}~\bibnamefont {Zsigmond}},\ }\href {\doibase 10.1364/OE.23.022108}
  {\bibfield  {journal} {\bibinfo  {journal} {Opt. Express}\ }\textbf {\bibinfo
  {volume} {23}},\ \bibinfo {pages} {22108} (\bibinfo {year}
  {2015})}\BibitemShut {NoStop}%
\bibitem [{\citenamefont {Ingleby}\ \emph {et~al.}(2017)\citenamefont
  {Ingleby}, \citenamefont {Griffin}, \citenamefont {Arnold}, \citenamefont
  {Chouliara},\ and\ \citenamefont {Riis}}]{Ingleby2017}%
  \BibitemOpen
  \bibfield  {author} {\bibinfo {author} {\bibfnamefont {S.~J.}\ \bibnamefont
  {Ingleby}}, \bibinfo {author} {\bibfnamefont {P.~F.}\ \bibnamefont
  {Griffin}}, \bibinfo {author} {\bibfnamefont {A.~S.}\ \bibnamefont {Arnold}},
  \bibinfo {author} {\bibfnamefont {M.}~\bibnamefont {Chouliara}}, \ and\
  \bibinfo {author} {\bibfnamefont {E.}~\bibnamefont {Riis}},\ }\href {\doibase
  10.1063/1.4980159} {\bibfield  {journal} {\bibinfo  {journal} {Rev. Sci.
  Instrum.}\ }\textbf {\bibinfo {volume} {88}},\ \bibinfo {pages} {043109}
  (\bibinfo {year} {2017})}\BibitemShut {NoStop}%
\bibitem [{\citenamefont {Colombo}\ \emph {et~al.}(2017)\citenamefont
  {Colombo}, \citenamefont {Dolgovskiy}, \citenamefont {Scholtes},
  \citenamefont {Gruji{\'{c}}}, \citenamefont {Lebedev},\ and\ \citenamefont
  {Weis}}]{Colombo2017}%
  \BibitemOpen
  \bibfield  {author} {\bibinfo {author} {\bibfnamefont {S.}~\bibnamefont
  {Colombo}}, \bibinfo {author} {\bibfnamefont {V.}~\bibnamefont {Dolgovskiy}},
  \bibinfo {author} {\bibfnamefont {T.}~\bibnamefont {Scholtes}}, \bibinfo
  {author} {\bibfnamefont {Z.~D.}\ \bibnamefont {Gruji{\'{c}}}}, \bibinfo
  {author} {\bibfnamefont {V.}~\bibnamefont {Lebedev}}, \ and\ \bibinfo
  {author} {\bibfnamefont {A.}~\bibnamefont {Weis}},\ }\href {\doibase
  10.1007/s00340-016-6604-8} {\bibfield  {journal} {\bibinfo  {journal} {Appl.
  Phys. B}\ } (\bibinfo {year} {2017}),\ 10.1007/s00340-016-6604-8}\BibitemShut
  {NoStop}%
\bibitem [{\citenamefont {Castagna}\ \emph {et~al.}(2009)\citenamefont
  {Castagna}, \citenamefont {Bison}, \citenamefont {{Di Domenico}},
  \citenamefont {Hofer}, \citenamefont {Knowles}, \citenamefont {Macchione},
  \citenamefont {Saudan},\ and\ \citenamefont {Weis}}]{Castagna2009}%
  \BibitemOpen
  \bibfield  {author} {\bibinfo {author} {\bibfnamefont {N.}~\bibnamefont
  {Castagna}}, \bibinfo {author} {\bibfnamefont {G.}~\bibnamefont {Bison}},
  \bibinfo {author} {\bibfnamefont {G.}~\bibnamefont {{Di Domenico}}}, \bibinfo
  {author} {\bibfnamefont {A.}~\bibnamefont {Hofer}}, \bibinfo {author}
  {\bibfnamefont {P.}~\bibnamefont {Knowles}}, \bibinfo {author} {\bibfnamefont
  {C.}~\bibnamefont {Macchione}}, \bibinfo {author} {\bibfnamefont
  {H.}~\bibnamefont {Saudan}}, \ and\ \bibinfo {author} {\bibfnamefont
  {A.}~\bibnamefont {Weis}},\ }\href {\doibase 10.1007/s00340-009-3464-5}
  {\bibfield  {journal} {\bibinfo  {journal} {Appl. Phys. B}\ }\textbf
  {\bibinfo {volume} {96}},\ \bibinfo {pages} {763} (\bibinfo {year} {2009})},\
  \Eprint {http://arxiv.org/abs/0812.4425} {arXiv:0812.4425} \BibitemShut
  {NoStop}%
\bibitem [{\citenamefont {Weis}\ \emph {et~al.}(2006)\citenamefont {Weis},
  \citenamefont {Bison},\ and\ \citenamefont {Pazgalev}}]{Weis2006}%
  \BibitemOpen
  \bibfield  {author} {\bibinfo {author} {\bibfnamefont {A.}~\bibnamefont
  {Weis}}, \bibinfo {author} {\bibfnamefont {G.}~\bibnamefont {Bison}}, \ and\
  \bibinfo {author} {\bibfnamefont {A.~S.}\ \bibnamefont {Pazgalev}},\ }\href
  {\doibase 10.1103/PhysRevA.74.033401} {\bibfield  {journal} {\bibinfo
  {journal} {Phys. Rev. A}\ }\textbf {\bibinfo {volume} {74}},\ \bibinfo
  {pages} {033401} (\bibinfo {year} {2006})},\ \Eprint
  {http://arxiv.org/abs/0605234} {arXiv:0605234 [physics]} \BibitemShut
  {NoStop}%
\bibitem [{\citenamefont {Bevilacqua}\ \emph {et~al.}(2014)\citenamefont
  {Bevilacqua}, \citenamefont {Breschi},\ and\ \citenamefont
  {Weis}}]{Bevilacqua2014}%
  \BibitemOpen
  \bibfield  {author} {\bibinfo {author} {\bibfnamefont {G.}~\bibnamefont
  {Bevilacqua}}, \bibinfo {author} {\bibfnamefont {E.}~\bibnamefont {Breschi}},
  \ and\ \bibinfo {author} {\bibfnamefont {A.}~\bibnamefont {Weis}},\ }\href
  {\doibase 10.1103/PhysRevA.89.033406} {\bibfield  {journal} {\bibinfo
  {journal} {Phys. Rev. A}\ }\textbf {\bibinfo {volume} {89}},\ \bibinfo
  {pages} {033406} (\bibinfo {year} {2014})}\BibitemShut {NoStop}%
\bibitem [{\citenamefont {Morrison}\ and\ \citenamefont
  {Parker}(1987)}]{Morrison1987}%
  \BibitemOpen
  \bibfield  {author} {\bibinfo {author} {\bibfnamefont {M.~A.}\ \bibnamefont
  {Morrison}}\ and\ \bibinfo {author} {\bibfnamefont {G.~A.}\ \bibnamefont
  {Parker}},\ }\href {\doibase 10.1071/PH870465} {\bibfield  {journal}
  {\bibinfo  {journal} {Australian Journal of Physics}\ }\textbf {\bibinfo
  {volume} {40}},\ \bibinfo {pages} {465} (\bibinfo {year} {1987})},\ \Eprint
  {http://arxiv.org/abs/arXiv:1011.1669v3} {arXiv:arXiv:1011.1669v3}
  \BibitemShut {NoStop}%
\end{thebibliography}%
\end{document}